\documentclass[a4paper,showpacs,prl,twocolumn]{revtex4} 
\usepackage{amsfonts}
\usepackage{amsmath}
\usepackage{amssymb}
\usepackage{graphicx}

\begin{document}

\title{Origin of the \textquotedblleft 0.25-anomaly\textquotedblright\ in
the nonlinear conductance of a quantum point contact}
\author{S. Ihnatsenka}
\affiliation{Department of Physics, Simon Fraser University, Burnaby, British Columbia V5A 1S6,
Canada}
\author{I. V. Zozoulenko}
\affiliation{Solid State Electronics, ITN, Link\"{o}ping University, 601 74, Norrk\"{o}%
ping, Sweden}
\date{\today }

\begin{abstract}
We calculate the non-linear conductance of a quantum point contact using the
non-equilibrium Greens function technique within the Hartree approximation of spinless
electrons. We quantitative reproduce the \textquotedblleft
0.25-anomaly\textquotedblright\ in the differential conductance (i.e. the lowest plateau
at $\sim 0.25-0.3\times 2e^2/h$) as well as an upward bending of higher conductance
half-integer plateaus seen in the experiments, and relate these features to the
non-linear screening and pinning effects.
\end{abstract}

\maketitle

\section{Introduction} A quantum point contact (QPC) is a narrow constriction of a width
comparable to the electron wavelength defined in a two-dimensional electron gas by means
of split-gate or etching technique. Due to quantization of the transverse motion
electrons can propagate only via allowed modes and the low-temperature linear-response
conductance of the QPC shows a step-like dependence on a gate voltage \cite{Wharam}. When
a bias voltage $V_{sd}$ is applied between the source and drain electrodes the integer
steps in the differential conductance $N\times 2e^{2}/h$ are
smoothed and gradually transformed into the half-integer plateaus $(N-\frac{1%
}{2})\times 2e^{2}/h$, where $N=1,2,3,...$ is a number of channels available
for propagation \cite%
{Glazman,Patel,MartinMoreno,Frost_Berggren,Kristensen,Cronenwett,Picciotto,Kothari,Chen}%
. Many features of the linear and nonlinear response of the QPC are by now well
understood. However, even after 20 years that have passed since the discovery of the
conductance quantization, some of important aspects of the QPC conductance are not
resolved yet and are still under discussions. One of the prominent examples (apart from
the famous \textquotedblleft 0.7-anomaly\textquotedblright\ \cite{0.7 anomaly}) is a
so-called \textquotedblleft 0.25-feature\textquotedblright\ in the non-equilibrium
differential conductance whose origin is under lively current debate \cite%
{Picciotto,Kothari,Chen}.

A theory of the non-equilibrium conductance of the QPC predicting the above
mentioned half-integer plateaus was developed by Glazman and Khaetskii \cite%
{Glazman}. The half-integer plateaus have subsequently been observed and
thoughtfully studied by a number of groups \cite%
{Patel,MartinMoreno,Frost_Berggren,Kristensen,Cronenwett,Picciotto,Kothari,Chen}%
. The theory of Glazman and Khaetskii \cite{Glazman} and a later more
refined approach by Frost \textit{et al.} \cite{Frost_Berggren} successfully
describe the QPC conductance in the regime when the differential conductance
$G_{d}\gtrsim 2e^{2}/h$. However, for $G_{d}\lesssim 2e^{2}/h$ instead of
the expected plateau at $(0.5)\times 2e^{2}/h$ practically all experiments
show a plateau at $(0.2-0.3)\times 2e^{2}/h$ (sometimes called as a
\textquotedblleft 0.25-feature\textquotedblright ) \cite%
{Patel,MartinMoreno,Frost_Berggren,Kristensen,Cronenwett,Picciotto,Kothari,Chen}%
. It has been recently argued that the \textquotedblleft
0.25-feature\textquotedblright\ corresponds to the fully spin polarized
current even at zero magnetic field \cite{Chen}. This conclusion implies far
reaching consequences for semiconductor spintronics as it opens up exciting
possibilities to generate spin polarized current simply by applying a
source-bias voltage to the quantum wire. However, alternative explanations
of the \textquotedblleft 0.25-feature\textquotedblright\ due to the
self-consistent electrostatics and non-linear screening of the lowest
\textit{spin-degenerate} subband have been advocated by other groups \cite%
{Picciotto,Kothari}. In particular, Kothari \textit{et al.} \cite{Kothari}
demonstrated that the experimental data are well-described by the analytical
models of Frost \textit{et al.} \cite{Frost_Berggren} with phenomenologically
introduced asymmetric voltage drop between the source and the drain.

A detailed understanding of the QPC conductance is of the prime importance
because the QPC represents the cornerstone of mesoscopic physics, and the
conductance quantization is a fundamental phenomenon of electron transport
in low-dimensional structures. The controversy concerning the origin of the
\textquotedblleft 0.25-feature\textquotedblright\ outlines the need for
microscopic modeling based on the self-consistent approaches to the electron
interaction and non-linear screening free from adjustable parameters. It
should be stressed that previous phenomenological approaches \cite%
{Glazman, MartinMoreno, Frost_Berggren, Kothari}, while providing an important insight for
interpretation of experiments, are not, however, able to uncover a microscopic origin of
the observed feature.

In this paper we present a model within the self-consistent Hartree approximation that
allows us to describe the nonlinear screening and evolution of the conductance plateaus
out of equilibrium and thus uncover underlying microscopic origin of the observed
features in the differential conductance. To solve the Schr\"{o}dinger equation we employ
a standard
non-equilibrium Green's function (NEGF) formalism \cite{Datta_book,Xue-Datta}%
. We demonstrate that for $G\lesssim 2e^{2}/h$ the differential conductance
exhibits $\sim (0.25-0.3)\times 2e^{2}/h$ plateau (as opposed to the $%
0.5\times 2e^{2}/h$ plateau predicted by the noninteracting theories \cite%
{Glazman,Frost_Berggren}). We also find that in the regime of $G\gtrsim
2e^{2}/h$ the nonlinear screening causes the half-integer plateaus to bend
upward as $V_{sd}$ increases. Note that this bending can be clearly seen in
all the reported experiments \cite%
{Patel,MartinMoreno,Frost_Berggren,Kristensen,Cronenwett,Picciotto,Kothari,Chen}%
, but, surprisingly enough, its presence passed without comments (except of
a brief discussion in \cite{MartinMoreno}). Our finding therefore strongly
indicates that \textquotedblleft 0.25-feature\textquotedblright\ is not
spin-related and is caused by the non-linear screening and related pinning
of spin-degenerate electrons in the QPC.

\section{Model} 
\begin{figure}[tbh!]
\includegraphics[keepaspectratio,width=\columnwidth]{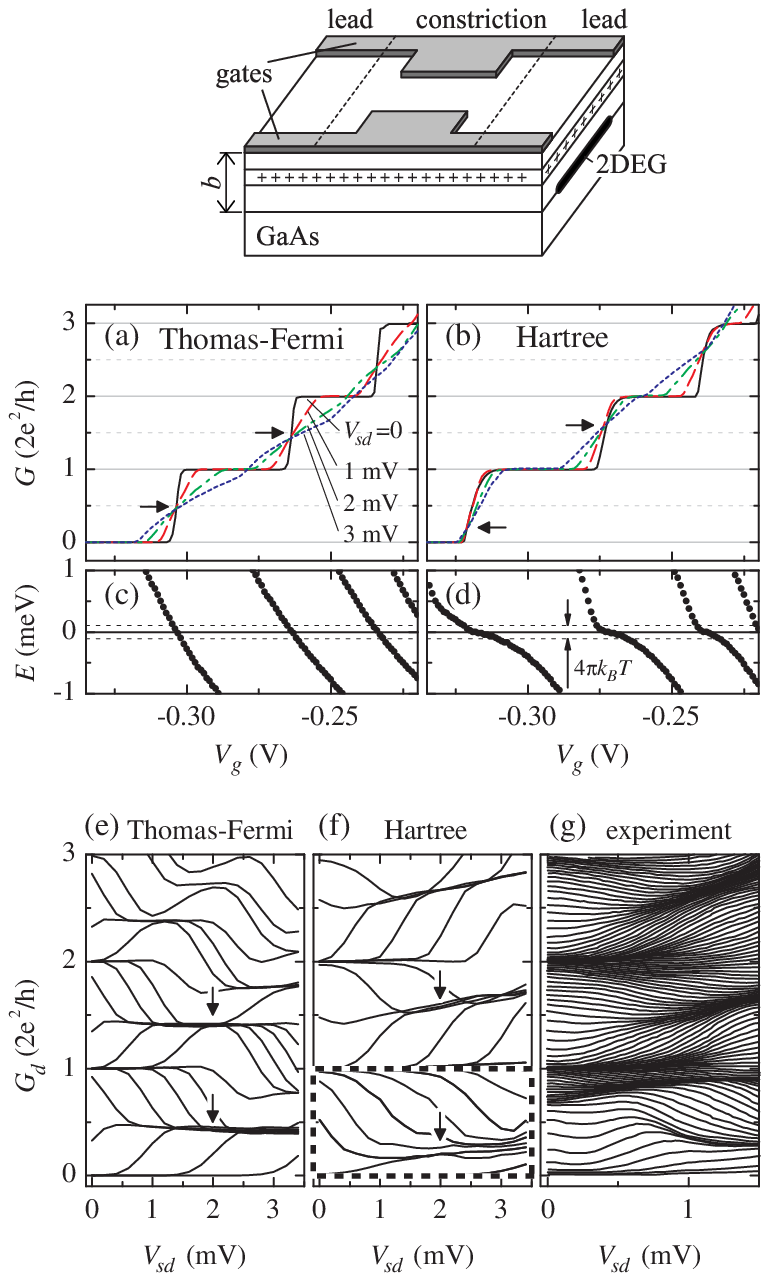}
\caption{(a),(b) Conductance $G$ of the QPC calculated within the TF and Hartree
approximations for different bias voltages $V_{sd}$. (c),(d) Resonant energy structure
(i.e. the LDOS integrated over the geometrical area of the QPC constriction) in
equilibrium, $V_{sd}=0$. (e),(f) The differential conductance $G_{d}$ calculated within
the TF and Hartree approximations. Traces are taken at different gate voltages with 5 mV
step (the dashed rectangular in (f) contains curves with 2.5 mV step);(g) shows the
experientially measured $G_d$ adapted from \protect\cite{Cronenwett}. The inset on the
top illustrates a geometry of the QPC defined in a GaAs heterostructure. A negative
voltage is applied to the top gates depleting the 2DEG residing on the distance $b=60$ nm
beneath the surface. The widths of the cap, donor, and spacer layers are 14, 36, and 10
nm, respectively; the donor concentration is $0.64\cdot 10^{24}$ m$^{3}$. The geometrical
width and length of the constriction are respectively 60 nm and 400nm. Temperature
$T=0.2$ K.} \label{fig:conductance}
\end{figure}

We consider a QPC defined by split gates in a GaAs heterostructure; see
Fig. \ref{fig:conductance}. The Hamiltonian of the whole system (the QPC plus the
semi-infinite leads) can be written in the
form $H(\mathbf{r})=-\frac{\hbar ^{2}}{2m^{\ast }}\nabla ^{2}+V_{eff}(%
\mathbf{r}),$ where $\mathbf{r}=(x,y)$, $m^{\ast }=0.067m_{e}$ is the GaAs
effective mass. The effective potential
\begin{equation}
V_{eff}(\mathbf{r})=V_{conf}(\mathbf{r})+V_{H}(\mathbf{r})+V_{bias}(\mathbf{r%
}),  \label{eq:Veff}
\end{equation}%
is the sum of the electrostatic confinement (including contributions from the top gates,
the donor layer, and the Schottky barrier), the Hartree and the bias potentials (see
\cite{opendot} for details). The Hartree potential is
written in a standard form \cite{Davies_book,opendot} $V_{H}(\mathbf{r})=%
\frac{e^{2}}{4\pi \varepsilon _{0}\varepsilon _{r}}\int d\mathbf{r}%
\,^{\prime }n(\mathbf{r}^{\prime })\left( \frac{1}{|\mathbf{r}-\mathbf{r}%
^{\prime }|}-\frac{1}{\sqrt{|\mathbf{r}-\mathbf{r}^{\prime }|^{2}+4b^{2}}}%
\right)$, where $n(\mathbf{r})$ is the electron density, $\varepsilon _{r}=12.9$ is the
dielectric constant of GaAs, and the second term describes
the mirror charges placed at the distance $b$ from the surface, Fig. \ref%
{fig:conductance}. The integration is performed over the whole device area
including the semi-infinite leads; e.g., the Coulomb interaction is included
both in the leads and in the QPC regions.

The Fermi energies $E_{F}$ in the left ($L$) and right ($R$) leads are
shifted by the applied source-drain voltage $V_{sd}$, $%
E_{F}^{L}=E_{F}^{R}+eV_{sd},$ while there is a linear ramp of $V_{bias}(%
\mathbf{r})$ over the device region \cite{Xue-Datta} (we set $E_{F}^{R}=0)$.
For a finite bias the electric current is calculated as \cite{Davies_book} $%
I=\frac{2e}{h}\int dE\;T(E)\left[ f_{L}^{FD}(E)-f_{R}^{FD}(E)\right] ,$ with
$T(E)$ being the transmission coefficient and $f_{L(R)}^{FD}(E)$ is the
Fermi-Dirac (FD) distribution in the left (right) leads. To calculate $T(E)$%
, the electron density and the local density of states (LDOS) we use the standard NEGF
method \cite{Datta_book,Xue-Datta} (see Appensix for the details of our calculations).
Having calculated the current $I$ we are in position to calculate the conductance
$G=I/V_{sd}$ and the differential conductance $G_{d}=dI/dV_{sd}$. The latter we compute
by increasing the bias voltage slightly and calculating the derivative $%
dI/dV_{sd}$ numerically.

To outline the role of quantum-mechanical effects in the electron-electron
interaction in the QPC, we also consider the Thomas-Fermi (TF) approximation
solving the standard TF equation to find the effective TF potential \cite%
{opendot} and calculating $G$ and $G_{d}$ for this potential using the NEGF.
This approximation does not capture quantum-mechanical quantization of
electron motion and, therefore, utilization of the TF approximation is
conceptually equivalent to a one-electron noninteracting approach.

\section{Results and discussion} Figure \ref{fig:conductance}(a),(b) shows the conductance $%
G$ of the QPC calculated within the TF and Hartree approximations for different
source-drain voltages $V_{sd}$. The parameters of the QPC are indicated in Fig.
\ref{fig:conductance} and are chosen close to those typically used in experiments (Note
that we performed calculations for shorter QPCs which show the same behavior). For zero
$V_{sd}$ the conductance shows well defined quantized plateaus for both TF and Hartree
approaches. The latter, however, predicts broader transition regions between the
plateaus. The reason is the energy level pinning effect\cite{opendot}. This is
illustrated in Fig. \ref{fig:conductance} (c), (d) that shows the resonant energy
structure inside the QPC constriction (i.e. the position of the peak in the local density
of states (LDOS) integrated over the geometrical area of the QPC constriction). In the TF
approach the resonant levels sweep past $E_{F}$ in a linear fashion. In contrast, the
Hartree calculations show pinning of the energy levels (corresponding to the
one-dimensional (1D) subbands in the narrowest part of the constriction) to $F_{F}$
within the energy window $\pm 2\pi k_{B}T$. Within this window the FD distribution
$0<f^{_{FD}}<1$, and thus the states are only partially filled. This leads to
metallic-like behavior when electrons can be easily rearranged to screen the external
electric field. (For influence of the pinning effect on equilibrium
transport in quantum dots see \cite{opendot}; see also \cite%
{Cambridge_pinning} for the experimental studies of the energy level pinning
in the QPC in the magnetic field).

Out of equilibrium, the energy window $eV_{sd}$ providing current carrying
states increases as the source-drain voltage grows, and the conductance
plateaus become smeared, see Fig. \ref{fig:conductance} (a), (b). The
plateaus in the conductance $G$ completely disappears when $eV_{sd}$ exceeds
the 1D subband energy separation inside the QPC constriction. At the same
time, half-integer plateaus $(N-\frac{1}{2})\times 2e^{2}/h$ appear in the
the differential conductance $G_{d}$, see Fig. \ref{fig:conductance} (e),
(f)). A comparison of the TF and Hartree results show two profound
differences between the calculated $G_{d}$. First, the lowest Hartree
plateau $N=1$ occurs at $G_{d}^{H}\approx 0.25-0.3\cdot 2e^{2}/h$ as opposed
to the $G_{d}^{TF}=0.5\times 2e^{2}/h$ plateau predicted by the TF
calculations. Second, all TF plateaus are flat and rather independent of $%
V_{sd}$, whereas all higher Hartree plateaus $N\geq 2$ are bent upward as $%
V_{sd}$ increases. Note that these two features of the calculated $G_{d}^{H}$
are clearly seen in all reported experiments \cite%
{Patel,MartinMoreno,Frost_Berggren,Kristensen,Cronenwett,Picciotto,Kothari,Chen} (see
Fig. \ref{fig:conductance} (g) for a representative example).

\begin{figure}[tbh]
\includegraphics[keepaspectratio,width=\columnwidth]{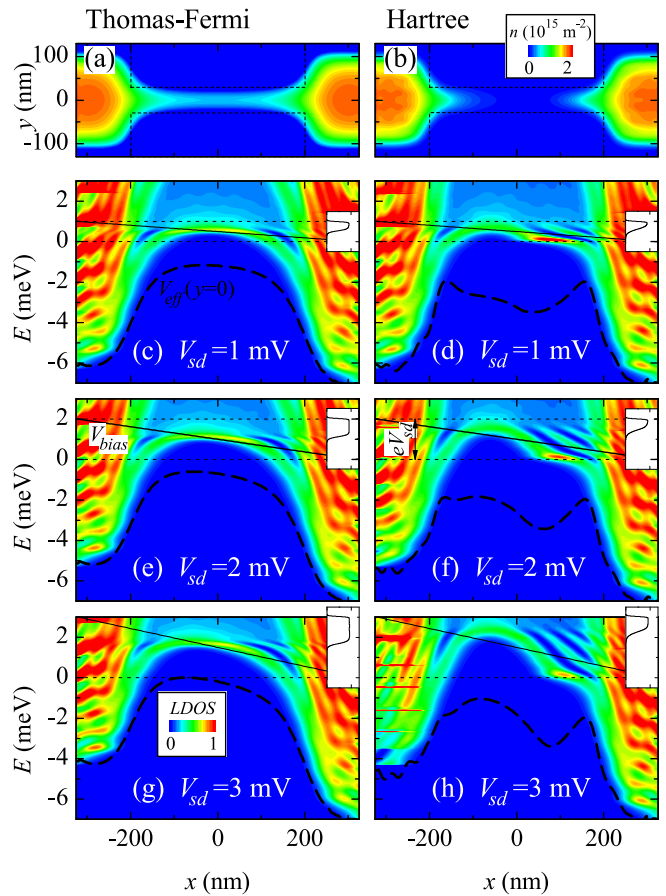}
\caption{The charge density and the LDOS of the QPC calculated in the
Thomas-Fermi and Hartree approximations (left and right panels respectively)
for the first half-integer plateau for different $V_{sd}$. The corresponding
gate voltages $V_g$ are marked by arrows in Figs. \protect\ref%
{fig:conductance}(a),(b),(e),(f). The effective potential Eq.
(\protect\ref{eq:Veff}) is plotted by the dashed lines. Solid slanted lines
denote the bias potential profile $V_{bias}$. Insets show the current
profiles, $T(E)\left[ f_{L}^{FD}(E)-f_{R}^{FD}(E)\right] $.}
\label{fig:DOS1}
\end{figure}
In order to understand the origin of the above features of the QPC nonlinear
conductance let us inspect the LDOS inside the QPC region. Let us first
concentrate at the first plateau in the differential conductance. Figures %
\ref{fig:DOS1} (c), (e), (g) show the evolution of LDOS as $V_{sd}$ is
increased calculated within the noninteracting TF approach. The enhanced
LDOS in the constriction corresponds to the position of the bottom of the
lowest propagating subband. In the TF approximation the effective
confinement potential is symmetrically distributed relative to $V_{bias}$
(that ramps linearly along the device). Because of this the 1D subband
touches $V_{bias}$ at the QPC center (at the energy $E=E_{F}^{L}-\frac{%
eV_{sd}}{2}$ ). As a result, the electrons injected from the left lead in
the upper half of the $eV_{sd}$ window ($E_{F}^{L}<E<E_{F}^{L}-\frac{eV_{sd}%
}{2}$) pass through the QPC with the unitary probability$.$ However, the
electrons in the lower half of the $eV_{sd}$ window experience a potential
barrier and hence are reflected back (see partial current profiles, $T(E)%
\left[ f_{L}^{FD}(E)-f_{R}^{FD}(E)\right] $, in small insets in Figs. \ref%
{fig:DOS1} (c), (e), (g)). Thus, the electrons injected from the left lead
give rise to the conductance of the half of the conductance unit, $%
G_{d}=0.5\times 2e^{2}/h.$ For electrons moving in the opposite direction,
from the drain to the source electrode, there is no available channel to
propagate and all of them are reflected.

A character of electron transport changes dramatically when interaction is included at
the quantum-mechanical level. Figures \ref{fig:DOS1} (d), (f), (h) show the LDOS inside
the constriction calculated within the Hartree approximation for the first half-integer
plateau where $G_{d}\approx 0.3\times 2e^{2}/h$. With one partially propagating mode the
electron density inside the constriction is low and the screening is rather week, and
hence the electron interaction strongly modifies the potential profile in comparison to
the symmetric TF distribution. The Coulomb charging pushes up the upper 1D subband inside
the QPC constriction to the top of the $eV_{sd}$ window near the source contact. (It is
interesting to note that the LDOS inside the QPC out of equilibrium resembles a
corresponding self-consistent LDOS profile of a resonant-tunneling diode \cite{Klimek}).
Thus, the 1D subbands becomes pinned to the top of the $eV_{sd}$ window and therefore
only a relatively narrow energy interval can supply current-carrying states that
pass through the QPC (see current profiles in the insets to Fig. \ref%
{fig:DOS1} (d),(f),(h)). Hence, the QPC conductance, $G_{d}\approx 0.3\times 2e^{2}/h$,
becomes smaller than a half of the conductance unit $G_0=2e^{2}/h$.  Our calculations
provide therefore a microscopic foundation of the phenomenological approaches that
describe the \textquotedblleft 0.25-anomaly\textquotedblright\ assuming a nonsymmetric
voltage drop inside the constriction\cite{Frost_Berggren,Kothari}.

\begin{figure}[tb]
\includegraphics[keepaspectratio,width=\columnwidth]{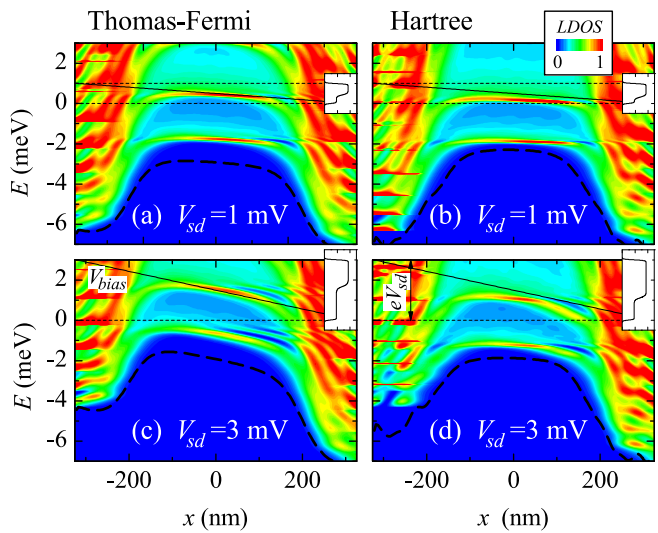}
\caption{TF and Hartree LDOS in the QPC for different $V_{sd}$. (The same as
in Fig. \protect\ref{fig:DOS1} but for the second half-integer plateau in
the differential conductance, see arrows at Figs. \protect\ref%
{fig:conductance}(a),(b),(e),(f).)}
\label{fig:DOS2}
\end{figure}

Let us now turn to higher half-integer plateaus. In this case there is at
least one propagating state inside the constriction, which, in turn, leads
to enhanced screening. Indeed, despite of the voltage drop between the left
and the right leads, the Hartree effective potentials $V_{eff}$ and the LDOS
are practically flat inside the QPC, see Figs. \ref{fig:DOS2} (b), (d). This
is in contrast to the corresponding TF results which do not account for
screening and thus follow the linear drop of $V_{bias}$, see Figs. \ref%
{fig:DOS2} (a), (c). Because of the enhanced screening, at the center of the QPC the
Hartree 1D subbands are situated lower than the corresponding TF subbands (i.e below
$E_{F}^{L}-\frac{eV_{sd}}{2}$ ), and hence the energy window providing the transmitted
states through the QPC exceeds the half of the available energy interval $eV_{sd}$ (see
the current profiles in the inset of Fig. \ref{fig:DOS2} (b), (d)). As a result, the QPC
conductance corresponding to the highest subband is larger than a half of the conductance
unit $G_0$. (Note that all lower subbands are fully occupied and thus contribute to one
conductance unit each). Thus, the enhanced screening, which becomes more pronounced as
$V_{sd}$ increases, is the reason for the upward bending of the higher half-integer
plateaus.

Finally, we stress that we utilized a model of \textit{spinless} electrons in the Hartree
approximation. The present approach can be easily extended to account for the spin effect
within the framework of the spin-density functional theory (SDFT). However, some previous
studies questioned the reliability of the SDFT for the system at hand because of the
self-interaction errors of the local spin density approximation \cite{07_QPC}%
. Hence, the definite answer about the role of the spin in the non-linear
conductance of the QPC might require approaches that go beyond the mean
field method used in the present study (e.g. quantum Monte-Carlo, etc.). At
the same time, an excellent quantitative agreement of our calculations with
the experimental results outlines the dominant role of the self-consistent
electrostatics and the nonlinear screening and strongly indicates that the
\textquotedblleft 0.25-feature\textquotedblright\ is not spin-related.

To conclude, using NEGF formalism within the Hartree model of spinless electrons we
reproduced quantitatively the observed features of the nonlinear QPC
conductance and provide microscopic interpretation of the \textquotedblleft
0.25-anomaly\textquotedblright\ as well as the upward bending of the higher
half-integer plateaus in terms of non-linear screening and pinning effect.

\acknowledgements This work has been supported by the Swedish Research Council (VR).

\appendix

\section{Appendix: Non-equilibrium Greens function (NEGF) technique for calculation of the
transmission coefficient of the QPC}

The central quantity in the NEGF is the lesser Green's function, $G^{<}$
\cite{Datta_book}. To calculate it one has to find first the retarded Green's function,
$G^{r}$,
\begin{equation}
\left( E-H(\mathbf{r})\right) G^{r}(\mathbf{r},\mathbf{r}^{\prime },E)=%
\mathbf{1},  \label{Green}
\end{equation}%
where $E$ is an electron energy and $\mathbf{1}$ is the unitary operator. This equation
can be reformulated using the so-called retarded self-energies of the leads, $\Sigma
_{R}^{r}$ and $\Sigma _{L}^{r}$,
\begin{equation}
\left( E-H_{0}-\Sigma _{R}^{r}(E)-\Sigma _{R}^{r}(E)\right) G^{r}(E)=\mathbf{%
1},
\end{equation}%
where $H_{0}$ is the Hamilton operator for the isolated scattering region (i.e. excluding
the leads). (For the sake of shortness we will not write explicitly a coordinate
dependence of $G^{r}$). $\Sigma _{R(L)}^{r}$ are functions with non-zero values only at
the boundaries with the semi-infinitive leads. Coupling the scattering region with leads
is described by the functions
\begin{subequations}
\begin{align}
i\Gamma _{R}(E)& =\Sigma _{R}^{r}(E)-\Sigma _{R}^{a}(E)=2i\:\mathrm{Im}\left[ \Sigma
_{R}^{r}(E)\right] , \\
i\Gamma _{L}(E)& =\Sigma _{L}^{r}(E)-\Sigma _{L}^{a}(E)=2i\:\mathrm{Im}\left[
\Sigma _{L}^{r}(E)\right] .
\end{align}%
The lesser Green's function in the scattering region is related to
electron flow from right and left reservoirs and is written as
\end{subequations}
\begin{eqnarray}
G^{<}(E)=&-&if_{R}^{FD}(E)\;G^{r}(E)\Gamma _{R}(E)G^{a}(E)  \notag \\
&-&if_{L}^{FD}(E)\;G^{r}(E)\Gamma _{L}(E)G^{a}(E), \label{eq:Glesser_general}
\end{eqnarray}%
where $f_{R(L)}^{FD}$ are the Fermi-Dirac functions in the right (left) lead. This
equation has to be used in non-equilibrium situations when $V_{sd}\neq 0$ and
$f_{R}^{FD}\neq f_{L}^{FD}$. In equilibrium, when the Fermi functions in both leads are
identical, Eq. \eqref{eq:Glesser_general} reduces to
\begin{equation}
G_{eq}^{<}(E)=2f_{R(L)}^{FD}(E)\;G^{r}(E).  \label{eq:Glesser_equilibrium}
\end{equation}%
It is also valid under a bias voltage at energies $E$ for which $%
f_{R}^{FD}=f_{L}^{FD}$ (in practice, $f_{R(L)}^{FD}=1$ for those energies).

In order to calculate the electron density we integrate over the electron energy $E$
\begin{equation}
n(\mathbf{r})=-\frac{1}{2\pi }\int dE\,\mathrm{Im}\left[ G^{<}\left( \mathbf{%
r},\mathbf{r},E\right) \right] .  \label{eq:density}
\end{equation}%
\begin{figure}[tb]
\includegraphics[scale=1]{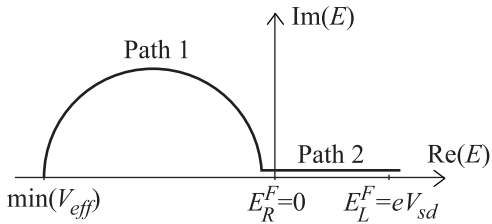}
\caption{ An integration path used in Eq. \eqref{eq:density}. Path 2 appears when the
bias voltage $V_{sd}$ applied.} \label{fig:structure}
\end{figure}
We use both Eqs. \eqref{eq:Glesser_general} and
\eqref{eq:Glesser_equilibrium} to perform this integration. $G_{eq}^{<}(E)$%
, Eq. \eqref{eq:Glesser_equilibrium}, is analytic in the upper half of the imaginary
plane whereas $G^{<}(E)$, Eq. \eqref{eq:Glesser_general}, has
poles below and above the real $E$-axis. Thus, for the energies when $%
f_{R(L)}^{FD}=1$ we can use $G_{eq}^{<}(E)$ for which we can transform the integration
path from the real axis to the complex plane \cite{Ihnatsenka}, see Fig.
\ref{fig:structure}, where $G_{eq}^{<}(E)$ is a smooth
function of energy. The rest of the integration (i.e. Path 2 in Fig. \ref%
{fig:structure}(b) where $f_{R(L)}^{FD}\neq 1)$, is close to the real axis and there Eq.
\eqref{eq:Glesser_general} is used. Along the Path 1 only
several integration points are needed because the rapid variations of $%
G_{eq}^{<}(E)$ are smeared out when the integration path is far from the real axis. This
is specially useful for the bound states, which give rise to sharp peaks near the real
axis. On the straight path along the real axis, one needs much more integration points
and for large source-drain voltage it becomes the most time consuming part of
computation.

Equations (\ref{Green})-(\ref{eq:density}) are solved self-consistently in an iterative
way until a converged solution for the electron density and potential (and hence for the
total Green's function) is obtained. Having calculated the total self-consistent Greens
functions, the transmission coefficient is calculated as \cite{Datta_book}
\begin{equation}
T(E)=\mathrm{Tr}\left[ \Gamma _{L}(E)G^{r}(E)\Gamma _{R}(E)G^{a}(E)\right] .
\label{eq:transmission}
\end{equation}
To speed up computation we employ the hybrid recursive method working with sin-Fourier
transformed Greens functions \cite{Zozoulenko_1996} and use the second Broyden method for
the iterative algorithm \cite{Broyden}.

\end{document}